\documentclass[prd,twocolumn,showpacs,showkeys]{revtex4}
\usepackage{graphicx}
\usepackage{amssymb}

\newcommand{\bear}{\begin{array}}  \newcommand{\eear}{\end{array}}
\newcommand{\bea}{\begin{eqnarray}}  \newcommand{\eea}{\end{eqnarray}}
\newcommand{\beq}{\begin{equation}}  \newcommand{\eeq}{\end{equation}}
\newcommand{\bef}{\begin{figure}}  \newcommand{\eef}{\end{figure}}
\newcommand{\bec}{\begin{center}}  \newcommand{\eec}{\end{center}}
\newcommand{\non}{\nonumber}

\newcommand{\ds}{\displaystyle}

\begin{document}

\title{Solving the discrepancy among the light elements abundances and WMAP}
\author{Kazuhide Ichikawa}
\email{kazuhide@icrr.u-tokyo.ac.jp}
\author{Masahiro Kawasaki}
\author{Fuminobu Takahashi}
\affiliation{Institute for Cosmic Ray Research, University of Tokyo, 5-1-5 Kashiwa-no-Ha, Kashiwa City, Chiba 277-8582, Japan}
\date{\today}

\begin{abstract}
Within the standard big bang nucleosynthesis (BBN) and cosmic
microwave background (CMB) framework, the baryon density measured by
the Wilkinson Microwave Anisotropy Probe (WMAP) or the primordial D
abundance is much higher than the one measured by the $^4$He or $^7$Li
abundances. To solve the discrepancy, we propose a scenario in which
additional baryons appear after BBN. We show that simply adding the
baryons can not be a solution but the existence of a large lepton
asymmetry before BBN makes the scenario successful. These extra
baryons and leptons, in addition to the initial baryons which exist
before the BBN, can be all produced from $Q$-balls.
\end{abstract}


\maketitle

\section{introduction} 
\label{sec:intro}

The baryon density is one of the most important cosmological
parameters. Especially, it is the only one input parameter for the
standard big bang nucleosynthesis (BBN) theory, which predicts the
abundances of light elements, D, $^4$He and $^7$Li. Meanwhile, for the
cosmic microwave background (CMB), it also plays an important role in
determining the shape of the acoustic peaks. Observations of the three
light elements and the CMB on the whole indicate equal amount of
baryons and make us confirm the validity of the standard cosmology.

Recently, following the precise measurement of the CMB by the
Wilkinson Microwave Anisotropy Probe (WMAP), its concordance with the
BBN and the light elements observations has been investigated in
Refs~\cite{Cyburt:2003fe,Coc:2003ce,Coc:2004ij,Cyburt:2004cq}. The
baryon density measured from the WMAP data is $\omega_b \equiv  
\Omega_b h^2 = 0.024 \pm 0.001$ (with the power-law $\Lambda$CDM
model)~\cite{Spergel:2003cb}, where $\Omega_b$ is the baryon energy
density divided by the critical energy density today and $h$ is the
Hubble constant in units of 100 km s$^{-1}$ Mpc$^{-1}$. The
uncertainty is very small because the WMAP has detected the first and
second peaks accurately in the temperature angular spectrum
\cite{Page:2003eu}. This corresponds to $\eta \equiv n_b/n_{\gamma} =
(6.6 \pm 0.3) \times 10^{-10}$, where $n_b$ and $n_{\gamma}$ are
baryon and photon number densities, via the relation $\eta =
\omega_b/(3.65 \times 10^7)$.
Refs~\cite{Cyburt:2003fe,Coc:2003ce,Coc:2004ij,Cyburt:2004cq} take
this well-determined WMAP $\omega_b$ as the BBN input\footnote{To be
more precise, they adopt $\eta = (6.14 \pm 0.25) \times 10^{-10}$
which is obtained from the running spectral index model value
$\omega_b = 0.0224 \pm 0.0009$ via the $\omega_b$-$\eta$ relation. Since this value is inferred from the combination of CMB, galaxy and Lyman $\alpha$ forest data, we adopt the value quoted in the text which is inferred using only the WMAP data.}
and calculate the light elements abundances and their theoretical
errors using improved evaluations of nuclear reaction rates and
uncertainties. The results are compared with the received measurements
of the primordial abundances of three light elements, D
(Ref.~\cite{Kirkman:2003uv}), $^4$He (Refs.~\cite{Fields1998} or
\cite{Izotov:2003xn}) and $^7$Li (Refs.~\cite{Ryan2000} or
\cite{Bonifacio:2002yx}). Although there are small differences
concerning their adopted reaction rates or observation data, their
conclusions agree: from the WMAP baryon density, the predicted
abundances are highly consistent with the observed D but not with
$^4$He or $^7$Li. They are produced more than observed. Especially, the $^7$Li-WMAP discrepancy is severer and it may require an explanation.

The most conservative and likely interpretation of such discrepancy is
that systematic errors in the primordial $^4$He and
$^7$Li measurements are underestimated in spite of the thorough analysis hitherto. 
Or, as Ref.~\cite{Coc:2003ce} has pointed out, it is possible that since the
cross section of the reaction $^7$Be(d,p)2$^4$He has not been measured
for the BBN energy range, it could reduce the $^7$Li yield to match
with the observation if the rate turns out to be hundreds times more
than the often neglected value obtained by the extrapolation.

If the discrepancy can not be attributed to systematic errors, we 
would have to invoke new physics to reduce their abundances. Actually, some approaches to reduce $^7$Li by astrophysical means, such as non-standard depletion mechanism inside stars, are discussed in Refs.~\cite{Cyburt:2003fe, Coc:2004ij,Romano:2003gv} and references therein.

What we seek in this paper is a cosmological solution. For a long time, not a few non-standard models are known to affect $^4$He abundance but none of them seem to be able to solve the discrepancy. For example, non-standard
expansion rate (extra relativistic degrees of freedom) and/or large
lepton asymmetry change $^4$He too much while adjusting $^7$Li to its observed value 
\cite{Barger:2003zg,Barger:2003rt,Steigman:2003gy}. A varying fine
structure constant (electromagnetic coupling constant) can relieve the
tension between either D and $^4$He or D and $^7$Li but not the three
elements together~\cite{Nollett:2002da}. At this time, only bolder
attempt such as to combine non-standard expansion rate and a varying
fine structure constant investigated by two of the authors in
Ref.~\cite{Ichikawa:2004ju} seems to be able to accommodate D, $^4$He
and $^7$Li simultaneously, but whether the introduction of these non-standard ingredients can fit the WMAP data is not fully checked yet \footnote{A varying deutron binding energy may have the capacity to render internal agreement between the light element abundances and with WMAP \cite{Dmitriev:2003qq}.}. 

In this paper, we investigate a solution which allows different amount of baryons
for the BBN and the CMB. Unfortunately this only reconciles the WMAP,
D and either $^4$He or $^7$Li. We choose to make $^7$Li consistent with the observation in
this way and consider a large lepton asymmetry in addition in order to
accommodate $^4$He too. One of the advantages of this scenario is that
both the additional baryons and lepton asymmetry can be produced from
$Q$-balls, which also generate the original baryons before the BBN.

In next section, we explain the discrepancy briefly and investigate
how the prediction of the primordial light element abundances is
affected by the additional baryons after the BBN. After we show that
is not enough to solve the discrepancy, we introduce the lepton
asymmetry and calculate how much extra baryons and lepton asymmetry
are needed to be a solution. In section~\ref{sec:qball}, we describe
how to generate required baryons and leptons. We conclude in
section~\ref{sec:conclusion}.

\section{A solution to the discrepancy}

In this section, we show that three elements abundances and the WMAP
data are not consistent within the standard framework of 
cosmology  but
they are reconciled with the appearance of additional baryons after
BBN and the existence of the large lepton asymmetry before the BBN. We
also compute how much of those are necessary to be a solution.

\subsection{The discrepancy among the light elements abundances and WMAP}

\begin{figure}
\includegraphics[width=7cm]{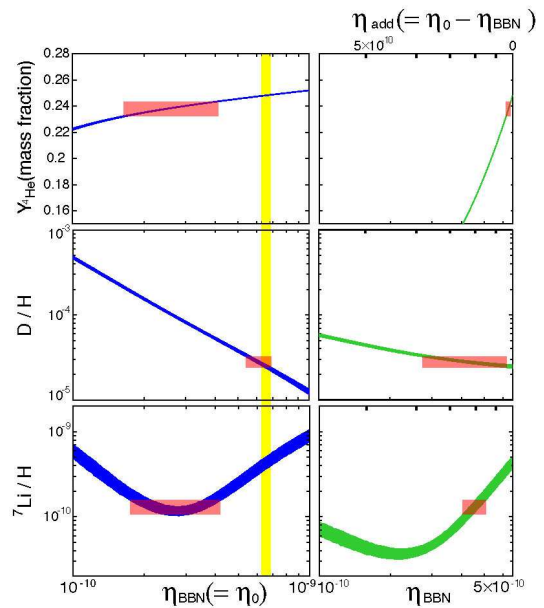}
\caption{
In the left column, familiar (namely, $\eta$ at present, $\eta_0$, is
same as the one at the BBN, $\eta_{\rm BBN}$.) standard BBN
calculations of $^4$He, D and $^7$Li abundances as functions of $\eta$
are expressed by three dark-shaded curves whose widths show
theoretical 1$\sigma$ uncertainties. The bar penetrating the figures
displays $\eta$ measured by the WMAP with 1$\sigma$ uncertainty. In
the right column, the light-shaded curves indicate the abundances
modified by the baryon appearance after the BBN. In this case, we
assume that $\eta$ is increased from $\eta_{\rm BBN}$ to be the WMAP
measured value, $\eta_0 = \eta_{\rm WMAP} = 6.6 \times 10^{-10}$. On
top of the figures, we mark the scale to tell the amount of added
$\eta$ after the BBN, where $\eta_0 = \eta_{\rm BBN}+\eta_{\rm
add}$. In both columns, the observational 1$\sigma$ uncertainties are
expressed by the vertical extension of the boxes, which are drawn to
overlap the theory curves so that their horizontal extension shows
allowed ranges of $\eta$.}
\label{fig:abundances}
\end{figure}

First of all, we summarize the measured baryon density from the BBN
and CMB in the left three panels in Fig~\ref{fig:abundances}. We
compute the theoretical abundances of $^4$He, D and $^7$Li and their
uncertainties by Monte Carlo simulations using the values obtained in
Ref.~\cite{Cyburt:2001pp} based on the reaction rates compiled in
Ref.~\cite{NACRE}. These are compared with the observations of three
light elements. We adopted the following data:
\begin{eqnarray}
{Y_{\rm ^4He}} &=& 0.238 \pm 0.002 \pm 0.005, \label{eq:obs_He_FO}\\
{\rm (D/H)} &=& 2.78^{+0.44}_{-0.38} \times 10^{-5}, \label{eq:obs_D}\\
{\rm (^7Li/H)} &=& 1.23^{+0.68}_{-0.32} \times 10^{-10}\ (95\%),
\label{eq:obs_Li_R}
\end{eqnarray}
where Eq.~(\ref{eq:obs_He_FO}) for $^4$He mass fraction is taken from
Ref.~\cite{Fields1998} (FO), Eq.~(\ref{eq:obs_D}) for D to H ratio in
numbers from Ref.~\cite{Kirkman:2003uv}, and Eq.~(\ref{eq:obs_Li_R})
for $^7$Li to H ratio in numbers from Ref.~\cite{Ryan2000} (R). In
Eq.~(\ref{eq:obs_He_FO}), the first uncertainty is statistical and the
second one is systematic, which are added in quadrature to be the
total observational uncertainty. The systematic errors are already included in the error bars of the other two elements. In Eq~(\ref{eq:obs_Li_R}), quoted
uncertainty is 95\% and we take its half value to be 1$\sigma$
uncertainty. As indicated by the three boxes in the figure, the $^4$He
and $^7$Li measurements indicate $\eta \approx (2 \sim 4) \times
10^{-10}$, but much higher value $\eta \approx (6 \sim 7) \times
10^{-10}$ is necessary to explain the D measurement.

On superposing the WMAP measured value \footnote{The WMAP group quotes
$\eta = (6.5^{+0.4}_{-0.3}) \times 10^{-10}$ as mean and 68\%
confidence range~\cite{Spergel:2003cb}. Although our adopted values
are slightly different since we derive them from $\omega_b$ as
explained in Sec.~\ref{sec:intro}, the difference scarcely affect our
results.},
\begin{eqnarray}
\eta = (6.6 \pm 0.3) \times 10^{-10}, 
\label{eq:eta_wmap}
\end{eqnarray}
the discrepancy seems to be severer. The range of
Eq.~(\ref{eq:eta_wmap}) is expressed by the narrow vertical bar in
Fig.~\ref{fig:abundances}, overlapping only with the $\eta$ range
deduced from D. The baryon density measured by the WMAP and the one by
either $^4$He or $^7$Li abundances are not very consistent.

We quantify the discrepancy by calculating $\chi^2$ as a function of $\eta$,
\begin{eqnarray}
\chi^2 = \sum_i \frac{(a_i^{th}(\eta) - a_i^{obs})^2}{(\sigma_i^{th})^2 + (\sigma_i^{obs})^2} + \frac{(\eta - \eta_{\rm WMAP})^2}{\sigma_{\rm WMAP}^2}, \label{eq:chi2}
\end{eqnarray}
where $a_i$ and $\sigma_i$ are respectively abundances and their 1$\sigma$ uncertainty of the element $i$. Their theoretical values are calculated from Monte Carlo simulations and observational values are those of Eqs.~(\ref{eq:obs_He_FO})$\sim$(\ref{eq:obs_Li_R}). For asymmetric errors, we adopt conservatively the larger one as 1$\sigma$ error (for $^7$Li, we divide the error in Eq.~(\ref{eq:obs_Li_R}) by two as explained above). We use the value of Eq.~(\ref{eq:eta_wmap}) for the second term. By taking the sum over three elements D, $^4$He and $^7$Li, we can investigate overall consistency and by choosing a particular element as $i$, we can study whether that element is consistent with the WMAP result.

\begin{figure}
\includegraphics[width=7cm]{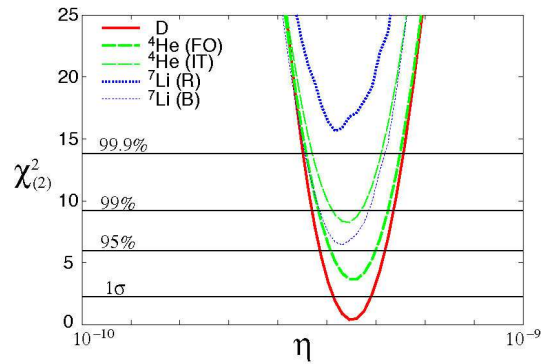}
\caption{$\chi^2_{(2)}$ calculated from a single element measurement and WMAP. The different types of lines use the data of D \cite{Kirkman:2003uv} (solid line), $^4$He of Fields and Olive (FO) \cite{Fields1998} (thick dashed line), $^4$He of Izotov and Thuan (IT) \cite{Izotov:2003xn} (thin solid line), $^7$Li of Ryan {\it et al.} (R) \cite{Ryan2000} (thick dotted line), and $^7$Li of Bonifacio {\it et al.} (B) \cite{Bonifacio:2002yx} (thin dotted line). The horizontal lines correspond to, from bottom to top, 1$\sigma$, 95\%, 99\% and 99.9\% confidence levels derived from two degree of freedom $\chi^2$.
}
\label{fig:chi2_bbn_wmap_1}
\end{figure}

\begin{figure}
\includegraphics[width=7cm]{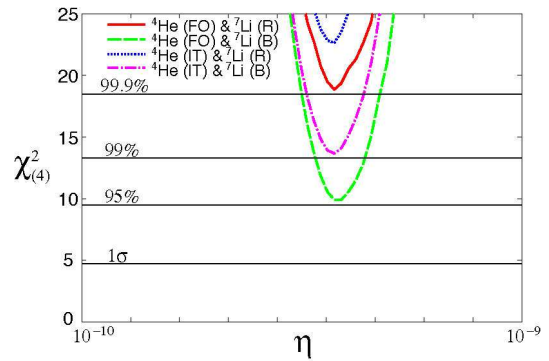}
\caption{$\chi^2_{(4)}$ calculated from three elements measurements and WMAP. Expressing each two measurements of $^4$He and $^7$Li by the abbreviations introduced in the caption to Fig.~\ref{fig:chi2_bbn_wmap_1}, the solid line uses FO for $^4$He and R for $^7$Li, the dashed line uses FO and B, the dotted line uses IT and R, and the dotted-dashed line uses IT and B. The horizontal lines correspond to, from bottom to top, 1$\sigma$, 95\%, 99\% and 99.9\% confidence levels derived from four degree of freedom $\chi^2$.  
}
\label{fig:chi2_bbn_wmap_2}
\end{figure}

The results are Figs.~\ref{fig:chi2_bbn_wmap_1} and \ref{fig:chi2_bbn_wmap_2}. Fig.~\ref{fig:chi2_bbn_wmap_1} shows the 2
two degree of freedom $\chi^2_{(2)}$ calculated from each element measurement and WMAP. Fig.~\ref{fig:chi2_bbn_wmap_2} shows the four degree of freedom $\chi^2_{(4)}$ from three elements measurements and WMAP. As expected from Fig.~\ref{fig:abundances}, Fig.~\ref{fig:chi2_bbn_wmap_1} shows that D is highly consistent with WMAP but not for $^4$He and $^7$Li. When we analyze with the combined data of three elements and WMAP, we see in Fig.~\ref{fig:chi2_bbn_wmap_2} that there is no baryon density range to explain the light elements abundances and the CMB. For details, $^4$He-WMAP discrepancy is not very severe but $^7$Li is definitely inconsistent with WMAP and this $^7$Li discrepancy mainly raises the overall $\chi^2$. Therefore, a task worth challenging would be to look for a cosmological model which can adjust $^7$Li abundance to the observed value while not amplifying the mild $^4$He-WMAP discrepancy. Of course, it is all the better to alleviate the tension of $^4$He in addition. 

We here comment on the other $^4$He and $^7$Li measurements. Ref.~\cite{Izotov:2003xn} (IT) reports ${Y_{\rm ^4He}} = 0.2421 \pm 0.0021$ and Ref.~\cite{Bonifacio:2002yx} (B) reports $\log [{\rm (^7Li/H)}] + 12 = 2.34 \pm 0.056 \pm 0.06$. The systematic effects are already included in the uncertainty in Ref.~\cite{Izotov:2003xn}'s $^4$He data. For Ref.~\cite{Bonifacio:2002yx}'s $^7$Li data, the statistical error and systematic error are added in quadrature so that  $[{\rm (^7Li/H)}]=2.19^{+0.46}_{-0.38} \times 10^{-10}$.The results using those data instead of Ref.~\cite{Fields1998}'s $^4$He and/or Ref.~\cite{Ryan2000}'s $^7$Li are also shown in Figs.~\ref{fig:chi2_bbn_wmap_1} and \ref{fig:chi2_bbn_wmap_2}. Ref.~\cite{Izotov:2003xn}'s $^4$He is less consistent with WMAP than Ref.~\cite{Fields1998}'s because of much less uncertainty in spite of the higher central value. It is tempting to address a strong $^4$He-WMAP discrepancy issue on the basis of Ref.~\cite{Izotov:2003xn}'s $^4$He but we adopt conservative Ref.~\cite{Fields1998}'s value to focus on $^7$Li discrepancy problem. For $^7$Li, we see that the use of Ref.~\cite{Bonifacio:2002yx}'s value relaxes the discrepancy with WMAP. We adopt Ref.~\cite{Ryan2000}'s measurement because they analyze the observations with lower and broader range of metallicity, but we note that considerable discrepancy still exists for Ref.~\cite{Bonifacio:2002yx}'s $^7$Li. 

\subsection{Effects of additional baryons after BBN}

Roughly speaking, the discrepancy exists because the WMAP data needs
more baryons than those required to account for the primordial light
elements abundances, especially $^4$He and $^7$Li. Then a naive
solution would be allowing to increase the baryons after the BBN to
the amount required to explain the WMAP data before the physics which
form the acoustic peaks takes place. 

However, such increase in the baryons (in the form of protons $i.e.$ H
nucleus) considerably affects the observation of the light elements
abundances because they are always measured in terms of the ratio to
H. Since the numerator does not change and the denominator increases,
the abundances decrease in general. To quantify this effect, let
$\eta_{\rm BBN}$ denote the baryon-to-photon ratio during the BBN and
let it increase by $\eta_{\rm add}$ (without entropy production) to be
$\eta_{\rm WMAP} \equiv 6.6 \times 10^{-10}$ after the BBN but well
before the matter-radiaion equality. They of course satisfy $\eta_{\rm
WMAP}=\eta_{\rm BBN}+\eta_{\rm add}$. For $^4$He, the effect is very
simple because its abundance is conventionally expressed by a mass
fraction: $Y_{\rm ^4He} \equiv 4n_{\rm ^4He}/n_b$, where $n_{\rm
^4He}$ is the number density of $^4$He. Noting that since entropy is
not produced, the photon number density scales as $a^{-3}$, where $a$
is the scale factor, the $^4$He mass fraction is modified from the
value when the BBN ends, $Y^{\rm (BBN)}_{\rm ^4He}$, to the one we
observe, $Y^{\rm (obs)}_{\rm ^4He}$, as
\begin{eqnarray}
{Y^{\rm (obs)}_{\rm ^4He}}={Y^{\rm (BBN)}_{\rm ^4He}}
\frac{\eta_{\rm BBN}}{\eta_{\rm  WMAP}}.
\label{eq:modHe4}
\end{eqnarray} 
For D and $^7$Li, since the abundance is expressed by the ratio to H
in numbers,
\begin{eqnarray}
{\rm (D/H)}_{obs} = {\rm (D/H)}_{\rm BBN}
 \frac{R_{\rm BBN}}{R_{\rm BBN}+\eta_{\rm add}},
 \label{eq:modD}
\end{eqnarray}
where $R_{\rm BBN}$ denotes the ratio of H to photon in numbers during
the BBN,
\begin{eqnarray}
R_{\rm BBN} \equiv \left[ \frac{n_{H}}{n_{\gamma}} \right]_{\rm BBN}
 = \eta_{\rm BBN} (1- Y^{\rm (BBN)}_{\rm ^4He}). 
\label{eq:ratio_H-gamma}
\end{eqnarray}
In Eq.~(\ref{eq:ratio_H-gamma}), we regard abundances of the light
elements other than H and $^4$He are negligibly small.

Using the Eqs.~(\ref{eq:modHe4}), (\ref{eq:modD}) and
(\ref{eq:ratio_H-gamma}), we can predict the observed abundances
modified by the baryon appearance after the BBN as shown in the right
three panels in Fig~\ref{fig:abundances}. Since we only consider the
increase in $\eta$, they are truncated at $\eta_{\rm WMAP}$ and
$\eta_{\rm add}$ increases to the left. Compared with the standard
case in the panels on the left hand side, the abundances become
smaller for greater $\eta_{\rm add}$ as expected and it is conspicuous
for $^4$He.

Similar to the standard BBN case, the comparison with the observations
is made by three boxes in the figure. We see D and either $^4$He or
$^7$Li can be reconciled with appropriate additional baryons but not
for three elements together.

To obtain a solution to reconcile three elements abundance
measurements simultaneously, we first choose to make D and $^7$Li
consistent by the adding baryons $\eta_{\rm add} \approx 1.5 \times
10^{-10}$. Then we introduce a (negative) lepton asymmetry before the
onset of the BBN in order to enhance $^4$He abundance (about 1.5
times) while retaining D and $^7$Li abundances. We next quantify how
much lepton asymmetry is needed.

\begin{figure}
\includegraphics[width=7cm]{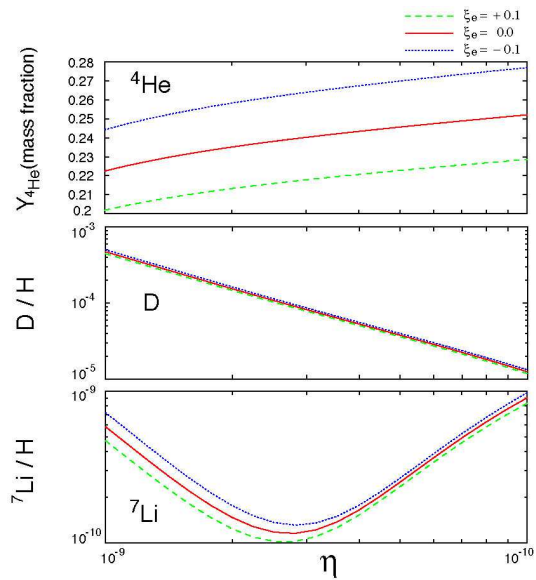}
\caption{
$\xi_e$-dependence of light element abundances. The cases for
$\xi_e=0$, $+0.1$, $-0.1$ are drawn with solid, dashed, and dotted
lines.}
\label{fig:xie_dependence}
\end{figure}

\subsection{Plus a lepton asymmetry before BBN}

The lepton asymmetry is parametrized by the degeneracy parameter
$\xi_e \equiv \mu_e/T$ where $\mu_e$ is the electron-type neutrino
chemical potential. Taking into account the tendency toward flavor
equilibration~\cite{Dolgov:2002ab}, we assume every flavor has the
same amount of asymmetry for concreteness.

Since we only consider $|\xi_e| \lesssim O(1)$, its contribution as
the extra relativistic degree of freedom is very small and the effect
on the nuclear beta equilibrium, $p+e^- \leftrightarrow n + \nu_e$
dominates. This property is important because it ensures that the baryon density measurement by CMB as Eq.~(\ref{eq:eta_wmap}) is not disrupted. When there is a negative lepton asymmetry ($\xi_e < 0$),
neutrinos are less than anti-neutrino and the equilibrium shifts to
increase $n$ relative to $p$. This is formulated as the following
equation. Since their number densities obey Boltzmann statistics, an
equilibrium number ratio at temperature $T$ is~\cite{EarlyUniverse}
\begin{eqnarray}
\frac{n_n}{n_p} = \exp \left[ -\frac{\Delta m}{T} - \xi_e \right],
\end{eqnarray}
where $\Delta m$ is the neutrino-proton mass difference and $\Delta m
\approx 1.29$MeV. From this expression it is easy to see $^4$He is
sensitive to non-zero $\xi_e$ because its abundance is approximately
written as
\begin{eqnarray}
Y_{^4{\rm He}} = \frac{2n_n}{n_n+n_p} \bigg|_f
 = \frac{2}{1+(n_p/n_n)_f} =\frac{2}{1+e^{\Delta m/T_f + \xi_e}},
 \label{eq:freezeout}
\end{eqnarray}
where the subscript $f$ means the values at the weak interaction
freeze-out and $T_f \approx 0.7$MeV, showing $^4$He mass fraction
depends exponentially on $\xi_e$. This is illustrated in the top panel
of Fig.~\ref{fig:xie_dependence}. The dependences of D and $^7$Li on
$\xi_e$ are also shown but they are very small. This is important
because we do not want to spoil the consistency between D and $^7$Li
by the introduction of the lepton asymmetry.

\begin{figure}
\includegraphics[width=7.5cm]{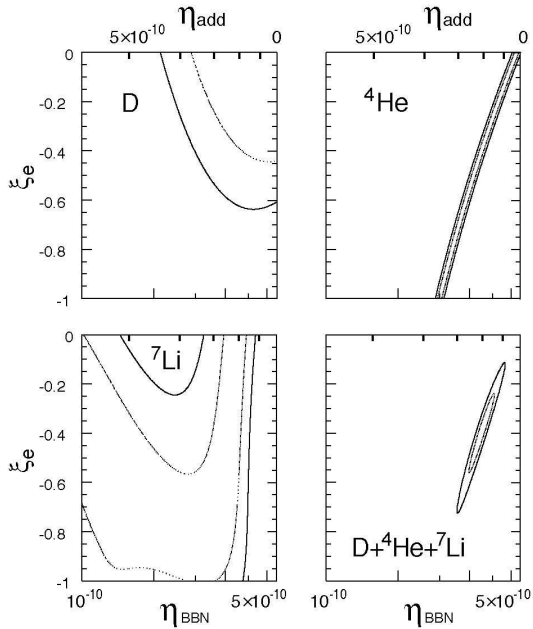}
\caption{
1$\sigma$ (dotted lines) and 95\% (solid lines) confidence levels in
the $\eta-\xi_e$ plane determined by each light elements observations
alone and by three elements combined.}
\label{fig:chi2}
\end{figure}

From Eq.~(\ref{eq:freezeout}), we estimate $\xi_e \sim -0.3$ is
necessary in order to make the $^4$He abundance larger and achieve the
scenario mentioned at the end of the previous subsection. To confirm
such an estimation, we search for the allowed parameter region on
$\eta_{\rm add}$-$\xi_e$ plane by calculating $\chi^2$ as a function
of those two parameters. Since we fix the baryon density after BBN at the WMAP central value, the $\chi^2$ here is calculated using three light elements data as Eq.~(\ref{eq:chi2}) with the last term omitted. The inclusion of the small uncertainty in WMAP baryon density would make the allowed region just a little larger. The results are displayed in Fig.~\ref{fig:chi2}. The
allowed regions determined from each element are understood from
Figs.~\ref{fig:abundances} and \ref{fig:xie_dependence}, and their
product set makes the region shown in the bottom-right panel, which
accommodates three elements. Especially, there is no solution for
$\xi_e = 0$ or $\eta_{\rm add} = 0$, and the best solution is located
at about $(\eta_{\rm add}, \xi_e) = (2 \times 10^{-10}, -0.4)$, as
expected during the argument thus far.

\section{$Q$-ball baryo- and leptogenesis}
\label{sec:qball}

In this section, we show a model that explains the large lepton
asymmetry, the initial baryon asymmetry before BBN, and additional
baryon appearance after the BBN all together. To accomplish this let
us consider the Affleck-Dine (AD) mechanism~\cite{Affleck:1984fy} and
the subsequent $Q$-ball formation~\cite{Kusenko:1997si,
Enqvist:1998en, Kasuya:1999wu} in the gauge-mediated supersymmetry
(SUSY) breaking model. Before turning to a close examination of the
model, it will be useful to summarize our basic strategy. The
mechanism generating large lepton asymmetry we adopt here is based on
Ref.~\cite{Kawasaki:2002hq}.  The AD mechanism can produce large (and
negative) lepton asymmetry by choosing an appropriate leptonic flat
direction such as $e_2L_1L_3$, where the subindices represent the
generation. After the flat direction starts oscillating, the AD field
experiences the spatial instability and deforms into nontopological
solitons, $Q$-balls ($L$-balls).  Then, almost all the produced lepton
numbers are absorbed into the $L$-balls~\cite{Kasuya:1999wu,
Kasuya:2001hg}. If the lifetime of such $L$-balls is longer than the
onset of electroweak phase transition but shorter than the epoch of
BBN, the large lepton asymmetry is protected from sphaleron
effects~\cite{Kuzmin:1985mm} and later released into the universe by
the decay of the $L$-balls. On the other hand, small (negative) lepton
numbers are evaporated from the $L$-balls due to thermal effects
before the electroweak phase transition, which are transformed into
small baryon asymmetry through the sphaleron effect. Thus generated
baryon number is the source for light element synthesis. Additional
baryon number after the BBN can also be explained by the use of
$Q$-balls. If the dimension of a flat direction is not so large, other
flat directions can obtain nonzero vevs simultaneously. For instance,
$udd$ direction is compatible with $eLL$ direction. The baryon
asymmetry generated by the $udd$ direction is also absorbed into
$Q$-balls ($B$-balls). If the $B$-balls decay after the relevant BBN
epoch, they can explain the late-time baryon appearance.  In order to
avoid dissociating the synthesized light elements with the decay
products, the mass per unit baryon charge of the $B$-ball must be
tacitly chosen to be slightly above the neutron or proton mass.  To
this end, we take $m_{3/2}=1$GeV in the following discussion (see
Eq.~(\ref{eq:mrq_for_new})).  It should be noted that the AD mechanism
and associated $Q$-ball formation can account for both the large
lepton asymmetry and the baryon asymmetry before and after the BBN.

\subsection{Affleck-Dine Mechanism and $Q$-balls}
\label{sec:review}

First off, let us review the AD mechanism and several properties of
$Q$-balls.  In the minimal supersymmetric standard model (MSSM) there
exist flat directions, along which there are no classical potentials
in the supersymmetric limit. Since flat directions consist of squarks
and/or sleptons, they carry baryon and/or lepton numbers, and can be
identified as the Affleck-Dine (AD) field.  For a definite discussion,
we adopt $eLL$ and $udd$ directions as the AD fields. It is of use to
parametrize the flat direction with a single complex scalar field
$\Phi$, so we express $eLL$ and $udd$ directions as $\Phi_L$ and
$\Phi_B$, respectively.

The flat directions are lifted by supersymmetry breaking
effects. In the gauge-mediated SUSY breaking model, the potential of a
flat direction is parabolic at the origin, and almost flat beyond the
messenger scale~\cite{Kusenko:1997si,Kasuya:2001hg,deGouvea:1997tn},
\begin{equation}
    V_{gauge} \sim \left\{ 
      \begin{array}{ll}
          m_{\phi}^2|\Phi|^2 & \quad (|\Phi| \ll M_S), \\
          \ds{M_F^4 \left(\log \frac{|\Phi|^2}{M_S^2} \right)^2}
          & \quad (|\Phi| \gg M_S), \\
      \end{array} \right.
\end{equation}
where $m_{\phi}$ is a soft breaking mass $\sim O(1 {\rm TeV})$, $M_F$ is the
SUSY breaking scale, and $M_{S}$ is the messenger mass scale.

Since gravity always exists, flat directions are also lifted by
gravity-mediated SUSY breaking effects~\cite{Enqvist:1997si},
\begin{equation}
    V_{grav} \simeq m_{3/2}^2 \left[ 1+K
      \log \left(\frac{|\Phi|^2}{M_*^2} \right)\right] |\Phi|^2,
\end{equation}
where $K$ is the numerical coefficient of the one-loop corrections and
$M_*$ is the gravitational scale ($\simeq 2.4 \times 10^{18}$
GeV). This term can be dominant only at high energy scales because of
small gravitino mass $\sim O(1\mbox{ GeV})$.

The non-renormalizable terms, if allowed by the symmetries of
Lagrangian, can exist and lift the flat directions;
\begin{equation}
    V_{NR} = \frac{|\Phi|^{2n-2}}{M^{2n-6}},
\end{equation}
where $M$ is a cut off scale of the non-renormalizable term. In our
scenario described in the next subsection, $\Phi_B$ is lifted by the
non-renormalizable potential, while $\Phi_L$ feels only the SUSY
breaking potentials.

The baryon and lepton number is usually created just after the AD
field starts coherent rotation in the potential, and its number
density $n_{B,L}$ is estimated as
\begin{equation}
\label{eq:number}
    n_{B,L}(t_{osc}) \simeq \varepsilon \omega \phi_{osc}^2,
\end{equation}
where $\varepsilon(\lesssim 1)$ is the ellipticity parameter, which
represents the strongness of the A term, and $\omega$ and $\phi_{osc}$
are the angular velocity and amplitude of the AD field at the
beginning of the oscillation (rotation) in its effective potential.

Actually, however, the AD field experiences spatial instabilities
during its coherent oscillation, and deforms into nontopological
solitons called
$Q$-balls~\cite{Kusenko:1997si,Enqvist:1998en,Kasuya:1999wu}.  When
the zero-temperature potential $V_{gauge}$ dominates at the onset of
coherent oscillation of the AD field, the gauge-mediation type
$Q$-balls are formed. Their mass $M_{Q}$ and size $R_Q$ are given
by~\cite{Dvali:1997qv}
\begin{equation}
    \label{eq:mass}
    M_Q \sim M_F Q^{3/4}, \qquad R_Q \sim M_F^{-1} Q^{1/4}.
\end{equation}
From the numerical simulations~\cite{Kasuya:1999wu,Kasuya:2001hg}, the
produced $Q$-balls absorb almost all the charges carried by the AD
field and the typical charge is estimated as~\cite{Kasuya:2001hg}
\begin{equation}
    Q \simeq \beta \left(\frac{\phi_{osc}}{M_F}\right)^4
\end{equation}
with $\beta \approx 6 \times 10^{-4}$.

There are also other cases where $V_{grav}$ dominates the potential at
the onset of coherent oscillation of the AD field. If $K$ is positive,
$Q$-balls do not form until the AD field leaves the $V_{grav}$
dominant region. Later it enters the $V_{gauge}$ dominant region and
experiences instabilities so that the gauge-mediation type $Q$-balls
are produced (delayed-type $Q$-balls)~\cite{Kasuya:2001hg}.  The
charge of the delayed-type $Q$-ball is
\begin{equation}
    \label{eq:delayq}
    Q \sim \beta \left(\frac{\phi_{eq}}{M_F}\right)^4
      \sim \beta \left(\frac{M_F}{m_{3/2}}\right)^4
\end{equation}
with $\phi_{eq} \sim M_F^2/m_{3/2}$. Here the subscript ``eq'' denotes
a value when the gauge- and the gravity-mediation potentials become
equal. Thus the delayed-type $Q$-balls are formed at $H_{eq} \sim
M_F^2/M_*$.

On the other hand, if the coefficient of the one-loop correction $K$
is negative, the gravity-mediation type $Q$-balls (``new'' type) are
produced~\cite{Kasuya:2000sc}. The typical value of their mass, size,
and charge are estimated as
\bea
\label{eq:mrq_for_new}
 M_Q &\sim& m_{3/2} Q, \qquad R_Q \sim (\sqrt{|K|}m_{3/2})^{-1},\non\\
 Q &\sim& \bar\beta \left(\frac{\phi_{osc}}{m_{3/2}}\right)^2
\eea
with $\bar\beta = 6.0 \times 10^{-3}$.

Finally let us mention the decay of $Q$-balls. In the case of
$L$-balls, they decay into leptons such as neutrinos via wino
exchanges. Also $B$-balls can decay into nucleons such as protons and
neutrons via gluino exchanges.  The decay rate of $Q$-balls is bounded
as~\cite{Cohen:1986ct}
\begin{equation}
    \label{eq:qdecay}
    \left|\frac{dQ}{dt}\right| \lesssim \frac{\omega^{3} A}{192 \pi^{2}},
\end{equation}
where $A$ is a surface area of the $Q$-ball. For $L$-balls, the decay
rate is estimated as a value of the order of the upper limit, while it
must be somewhat suppressed for $B$-balls, if the mass per unit baryon
number, $M_Q/Q$, is close to the nucleon mass $\sim 1$GeV.

\subsection{$B$- and $L$-balls}
\label{sec:$bl-ball}

Now we detail the scenario. In order to generate $O(1)$ lepton
asymmetry, the AD field responsible for large lepton asymmetry should
start to oscillate from the gravitational scale, {\it i.e.},
$\phi_{L,osc} = M_*$, which leads to the formation of  delayed-type
$L$-balls.  Meanwhile the value of $\phi_B$ at the onset of the
oscillation must be suppressed due to the existence of the
non-renormalizable term. Also $\phi_B$ is assumed to form new type
$B$-balls, which decay after the BBN.  Let us see if this setup can be
naturally realized step by step.

First of all, of the two flat directions, only $\phi_B$ should be
lifted by the non-renormalizable term, which can be explained by the
$R$-symmetery.  For instance, assigning $R$-charges $\frac{2}{3}$ to
$e$,$L$,$H_u$, and $H_d$, $\frac{1}{3}$ to $u$ and $d$, and $1$ to
$Q$, non-renormalizable terms for $eLL$ are forbidden, while the
following superpotential can lift the $udd$ direction;
\bea
W^{(udd)}_{NR} &=&  \frac{9(udd)^2}{2M^3} = \frac{\Phi_B^6}{6M^3},\non\\
V^{(udd)}_{NR} &=&   \frac{|\Phi_B|^{10}}{M^6}.
\eea
During inflation, $\phi_B$ thus sits at the minimum $\sim (H
M^3)^{1/4}$ determined by balancing the non-renormalizable term and
the negative Hubble-induced mass term.

Second, the abundance of the late-time baryon appearance must be of
the order of $\eta_{\rm add} \sim O(10^{-10})$. According to the
result of Ref.~\cite{Kawasaki:2002hq}, both the large lepton asymmery,
$|\xi_e| \sim O(0.1)$, and the baryon asymmetry necessary for the BBN,
$\eta_{\rm BBN} \sim O(10^{-10})$, can be successfully generated.\footnote{
To obtain positive baryon asymmetry, the total lepton asymmetry must
be negative.  The aim of Ref.~\cite{Kawasaki:2002hq} was to produce
``positive" lepton asymmetry of electron type, with the total lepton
asymmetry being negative. Here what we want is $\xi_e \sim O(-0.1)
<0$, therefore the flavor equilibration due to neutrino
oscillation~\cite{Dolgov:2002ab} does not spoil our scenario. However,
for recent work of suppressing such flavor equilibration, see
Ref.~\cite{Dolgov:2004jw}.
} Therefore it is enough to estimate the relative abundance of the
additional baryon number to the lepton asymmetry.  Using
Eq.~(\ref{eq:number}), the baryon number of the $udd$ direction is
suppressed by $(\phi_B/\phi_L)^2$, which is calculated as
\beq
\left(\frac{\phi_B}{\phi_L}\right)^2 
\sim \left(\frac{(m_{3/2} M^3)^{1/4}}{M_*}\right)^2
\sim 10^{-9},
\eeq
where we assumed $M\sim M_*$ and $m_{3/2} \sim 1$GeV. Thus we can naturally
obtain the desired abundance, $\eta_{\rm add} \sim O(10^{-10})$.

Thirdly, the sign of $K$ should be positive for $\Phi_L$, while
negative for $\Phi_B$. Unless the flat direction includes the third
generation superfields, the sign of $K$ tends to be negative due to
the contribution from the gauge interactions.  This is because the
contribution to $K$ from Yukawa (gauge) interactions is positive
(negative).  Note that the Yukawa coupling not only for top but also
for bottom and tau particles can be $O(1)$ for large $\tan
\beta$. Therefore, if we choose the flat directions as {\it e.g.},
$e_2L_1L_3$ and $u_1d_2d_1$, the sign of $K$ for the two directions is the
desired one.

Lastly, we need to show that the new type $B$-balls decay after the
relevant BBN epoch.  The decay rate of the $B$-ball is
\beq
\Gamma_Q \equiv \frac{1}{Q} \left|\frac{dQ}{dt}\right| = \frac{m_{3/2}^{5/2}}{48 \pi \bar\beta |K| M^{3/2}},
\eeq
where we used Eq.~(\ref{eq:qdecay}). Simply equating this quantity
with the Hubble parameter, we have the relatively high decay
temperature,
\beq
T_d \simeq 0.08\, {\rm MeV} \left(\frac{|K|}{0.1}\right)^{1/2} \left(\frac{m_{3/2}}{1{\rm GeV}}\right)^{5/4}
\left(\frac{M}{M_*}\right)^{-3/4},
\eeq
around which the light elements are being synthesized. However, as
noted above, the decay rate is suppressed if the mass per unit baryon
charge is very close to the nucleon mass, leading to smaller decay
temperature. Therefore we expect the new type $B$-balls decay after
the light element synthesis ceases to proceed.

To sum up the major characteristics of our scenario, the $Q$-balls
protect not only lepton asymmetry from the sphaleron effects, but also
baryon asymmetry from the BBN processes, which leads to the late-time
baryon appearance.

\section{conclusion} 
\label{sec:conclusion}
Concerning recent cosmological observations, there seems to be conflicting
measurements of the baryons in the universe, namely, the baryon density 
deduced from the CMB observation by the WMAP or the primordial D abundance
is much higher than the one derived from $^4$He or $^7$Li abundances. 
In this paper, we have proposed a scenario to reconcile such inconsistency by adding
baryons after the BBN and assuming a lepton asymmetry before the BBN. The introduction of the additional baryons leads to success in explaining the observed low abundance of $^7$Li without recourse to special models of stellar depletion. 
Also it was shown that the scenario can be naturally implemented in the
AD mechanism, where $Q$-balls play an essential role in preserving extra lepton
and baryon asymmetries. We should note that  this is the first cosmological scenario
that can completely remove the long-standing tension among the three light elements
and CMB. According to our result, amazingly, about one third of the baryon we observe 
in the present universe might have been ``sterile" during the BBN epoch. 
Such illuminating history of the universe might be confirmed through further
observational results with better accuracy in the future.

\section*{ACKNOWLEDGEMENTS}
This work was partially supported by the JSPS Grant-in-Aid for Scientific Research No.\ 10975 (F. T.).

 \end{document}